\documentstyle[12pt,psfig]{article}
\begin{document}
\hfill{\small ITP-SB-94-06}
\vglue 0.1cm
\vskip 1.5cm
\centerline{ {\large\bf Invariant mass distributions for heavy
quark-antiquark} }
\centerline{ {\large\bf pairs in deep inelastic electroproduction} }
\vskip 0.4cm
\centerline {\bf B. W. Harris and J. Smith }
\vskip 0.4cm
\centerline{\small \em Institute for Theoretical Physics,}
\centerline{\small \em State University of New York at Stony Brook,}
\centerline{\small \em Stony Brook, NY 11794-3840, USA}
\vskip 0.5cm

\centerline{January 1995}
\vskip 1.0cm
\centerline{Submitted to Physics Letters B}
\vskip 1.0cm
\centerline{\bf Abstract}

\vskip 1.0cm

We have completed the ${\cal O}(\alpha_s)$ QCD corrections to exclusive
heavy quark-antiquark distributions in deep inelastic electroproduction
and present here the differential distributions in the masses of
charm-anticharm and bottom-antibottom pairs at the electron-proton
collider HERA.

\vfill
\newpage
Order $\alpha_s$ QCD corrections to the structure functions for
single particle inclusive deep
inelastic electro-production of heavy quarks were recently published
in \cite{LRSvN1}.  The reaction
$e^{-}(l_1) + P(p) \rightarrow  e^{-}(l_2) + Q(p_1) + \bar Q(p_2) + X$ is
dominated by the virtual photon mediated reaction when
$-q^2 = -(l_1-l_2)^2 \ll M^2_Z$,
and the heavy quark differential production cross section
can be written as
%---(1)
\begin{equation}
\frac{d^2\sigma}{dxdy} = \frac{2\pi\alpha^2}{Q^4} S
[\{1+(1-y)^2\}F_2(x,Q^2,m^2)-y^2F_L(x,Q^2,m^2)] \, ,
\end{equation}
after integration over the azimuthal angle between the plane containing
the incoming and outgoing electron and the plane containing the incoming
proton and outgoing heavy quark.
The square of the center of momentum energy of the electron-proton
system is denoted by $S$, and the variables $x$ and $y$ are defined as
$x = Q^2/2p\cdot q$ and $y = p\cdot q / p\cdot l_1$
with $-q^2 = Q^2 = xyS $.  The heavy quark structure functions
$F_2(x,Q^2,m^2)$ and $F_L(x,Q^2,m^2)$ are functions of the heavy
quark mass $m$.  We assume that the heavy quark production is extrinsic so
that $F_2$ and $F_L$ can be calculated from an analysis of the virtual photon
induced reaction
$ \gamma^\ast(q) + P(p) \rightarrow  Q(p_1) + \bar Q(p_2) + X$
and its corresponding parton analogue
$ \gamma^\ast(q) + a_1(k_1) \rightarrow  Q(p_1) + \bar Q(p_2) + a_2(k_2)$,
where $a_1$ and $a_2$ are zero-mass gluons $g$ or light mass (anti) quarks
$(\bar{q}) \, q$ as
opposed to the massive (anti) quarks $(\bar Q) \, Q$.
The result is that the structure functions can be obtained from the
partonic results via the formula
%----(2)
\begin{eqnarray}
F_{k}(x,Q^2,m^2) &=&
\frac{Q^2 \alpha_s(\mu^2)}{4\pi^2 m^2}
\int_{\xi_{\rm min}}^1 \frac{d\xi}{\xi}  \Big[ \,e_H^2 f_g(\xi,\mu^2)
 c^{(0)}_{k,g} \,\Big] \nonumber \\&&
+\frac{Q^2 \alpha_s^2(\mu^2)}{\pi m^2}
\int_{\xi_{\rm min}}^1 \frac{d\xi}{\xi} \left\{ \,e_H^2 f_g(\xi,\mu^2)
\left( c^{(1)}_{k,g} + \bar c^{(1)}_{k,g} \ln \frac{\mu^2}{m^2} \right) \right.
\nonumber \\ &&
+\sum_{i=q,\bar q} f_i(\xi,\mu^2) \left. \left[ e_H^2
\left( c^{(1)}_{k,i} + \bar c^{(1)}_{k,i} \ln \frac{\mu^2}{m^2} \right)
+ e^2_i \, d^{(1)}_{k,i} + e_i\, e_H \, o^{(1)}_{k,i} \, \right]
\right\} \, , \nonumber \\ &&
\end{eqnarray}
where $(k = 2,L)$.
The lower boundary on the integration is given by
$\xi_{\rm min} = x(4m^2+Q^2)/Q^2$.
Further $f_i(\xi,\mu^2)\,, (i=g,q,\bar q)$ denote
the parton momentum distributions in the proton, and $\mu$ stands for the
mass factorization scale which has been put equal to the
renormalization scale in the running coupling constant.  Finally,
$c^{(l)}_{k,i}$ and $\bar c^{(1)}_{k,i}\,,
(i=g\,,q\,,\bar q\,;l=0,1)$,  and $d^{(1)}_{k,i}$ and
$o^{(1)}_{k,i}$, $(i=q\,,\bar q)$
are scale independent parton coefficient functions
which were first calculated in \cite{LRSvN1}.
In Eq. (2) we made a distinction between the coefficient functions with
respect to their origin.
The coefficient functions indicated by
$c^{(l)}_{k,i}$ and $ \bar c^{(1)}_{k,i}$ originate from the partonic
subprocesses where the virtual photon is coupled to the heavy quark hence
the factor of $e_H^2$.
The quantity $d^{(1)}_{k,i}$ comes from the subprocess where the virtual
photon interacts with the light quark so it is proportional to $e_i^2$.
The quantity $o^{(1)}_{k,i}$ comes from the interference between the
above processes and hence has a factor $e_H e_i$ with all charges
in units of $e$.
Note that terms proportional to $e_H e_i$ appear in the
photon-parton differential distributions even though they integrate
to zero in the total partonic cross section.
Furthermore we have isolated the factorization scale dependent term
containing $\ln(\mu^2/m^2)$. The functions multiplied by this term, which are
indicated by a bar, are called mass factorization parts.
Note than Eq. (2) only holds for $Q^2>0$.  In the photo-production limit
there are additional terms involving the parton densities in the photon.

The previous treatment of the ${\cal O}(\alpha_s)$ corrections
yielded results for the inclusive distributions for
heavy quarks in the virtual photon induced reaction
$ \gamma^\ast + P(p) \rightarrow  Q(p_1)(\bar Q(p_2)) + X$,
i.e., the differentials $dF_k(x,Q^2,m^2,p_t)/dp_t$
and $dF_k(x,Q^2,m^2,y)/dy$.
The corrections to the heavy quark
inclusive $p_t$ and $y$ distributions at fixed points
in $Q^2$ and $x$ were published in \cite{LRSvN2}.
Event rates for regions of the $x$ and $Q^2$ plane have been presented in
\cite{RSvN}.

In this paper we report on the results of a
calculation of the ${\cal O}(\alpha_s)$ corrections for heavy
quark {\em exclusive} distributions at fixed $Q^2$ and $x$,
which allows us to study all correlations between the outgoing
particles in the virtual photon initiated reaction
$\gamma^{\ast}(q) + P(p) \rightarrow Q(p_1) + \overline{Q}(p_2) + X(k_2)$
with $X=0$ or $1$ jet (massless parton).
This information is of immediate interest to the
experimenters working with the H1 and ZEUS collaborations at
the electron-proton collider HERA.
We therefore present the effects of the QCD corrections to
invariant mass distributions of a heavy quark-antiquark pair, for
$8.5 ({\rm GeV/c})^2 \leq Q^2 \leq 50 ({\rm GeV/c})^2$
and
$4.2 \times 10^{-4} \leq x \leq 2.7 \times 10^{-3}$.
The $(L,2)$ photon components are
treated separately.  As input we use the
latest CTEQ parton densities \cite{CTEQ}, which fit the
newly released HERA data \cite{HERA}.

Our new analysis of exclusive heavy quark
deep inelastic electro-production at HERA extends the
existing studies of inclusive QCD corrections in the virtual photon
channel \cite{LRSvN1}, \cite{LRSvN2}, \cite{RSvN},
inclusive QCD corrections in the real photon
$(q^2 = 0)$ channel \cite{EN}, \cite{SvN}, and exclusive QCD corrections
in the real photon channel \cite{frix}, allowing for an extensive
comparison with present and future experimental data.  Heavy quark
electro-production is expected to play an important role in the
determination of the gluon distribution
function in the proton at small $x$.
A knowledge of the production cross sections and
distributions for charm and bottom quarks is also
relevant in the study of
the CKM matrix elements through the rare decays of $D-$ and $B-$
mesons and the analysis of
$D\overline{D}$ and $B\overline{B}$ mixing \cite{Sch}.

In our exclusive computation we use the same techniques as the
authors of \cite{frix} for computing heavy-quark correlations in
photo-production and hadro-production.
These are based on the replacement of divergent terms in
the squared matrix elements by generalized plus distributions.
The divergent terms appear when
the propagators diverge in regions of phase space where
the outgoing parton is soft and/or collinear to the propagating particle.
The replacement of the divergent terms by generalized plus
distributions allows one to isolate the soft and collinear
poles within the framework of dimensional regularization, without
having to calculate all the phase space integrals
in a spacetime dimension $n\ne 4$ as usually required in
a traditional inclusive computation.
The resulting expressions for the squared matrix elements appear in a
factorized form where poles in $n-4$ multiply splitting functions and lower
order squared matrix elements.
The cancelation of singularities is then performed using the
factorization theorem \cite{CSS}. Since the final result
is in four-dimensional space time, we can compute all
relevant phase space integrations using standard Monte Carlo
integration techniques and produce histograms
for exclusive, semi-inclusive, or inclusive  quantities
related to any of the outgoing particles. We therefore have
a new calculation of the scale independent coefficient functions
$c^{(l)}_{k,i} \, , \, \bar c^{(1)}_{k,i} \, , \, d^{(1)}_{k,i}$
and $o^{(1)}_{k,i}$.
We checked the $\eta=s/4m^2-1$ and $\xi=Q^2/m^2$ dependence of
the scale independent coefficient functions
against the results in \cite{LRSvN1}. The
analogous results for the inclusive distributions,
$dF_{k}/dp_{t}$ and $dF_{k}/dy$,
were also checked against the results in \cite{LRSvN2}.
Additional distributions and correlations along with details of the
calculation will be presented in a more complete article \cite{HS2}.

Folding the parton densities in the proton with our new scale independent
coefficient functions as dictated by Eq. (2), we present results for
the differential structure
functions in the invariant mass of the heavy quark-antiquark pair, which
we will call $M$. Thus we give plots of
$dF_2(x,Q^2,m^2,M)/dM$ and
$dF_L(x,Q^2,m^2,M)/dM$ at fixed $x$ and $Q^2$.
We use $m=m_c=1.5 \, {\rm GeV/c^2}$ for charm production and
$m=m_b=4.75 \, {\rm GeV/c^2}$ for bottom production.  We choose
the factorization (renormalization) scale as
$\mu^2=Q^2+4(m_c^2+(P_{t_c}+P_{t_{\bar{c}}})^2/4)$ for charm production and
$\mu^2=Q^2+m_b^2+(P_{t_b}+P_{t_{\bar{b}}})^2/4$ for bottom production.
As mentioned earlier we use the CTEQ3M parton densities \cite{CTEQ}
in the $\overline{\rm MS}$ scheme and the two loop $\alpha_s$ with
$\Lambda_{4} = 0.239 \, {\rm GeV}$ for charm and
$\Lambda_{5} = 0.158 \, {\rm GeV}$ for bottom.

Tables 1 and 2 show the variation of $F_2(x,Q^2,m^2)$ and
$F_L(x,Q^2,m^2)$ with the renormalization scale
for charm production, and Tables 3 and 4 show the variation
for bottom  production.  For charm production typical variations
from the central value are less than $15\%$ and for bottom
production they are less than $6\%$.

Figures 1 and 2 show the distributions $dF_2(x,Q^2,m_c^2,M)/dM$
and $dF_L(x,Q^2,m_c^2,M)/dM$ for charm production
at $x=8.5 \times 10^{-4}$ while varying $Q^2$.
Figure 3 and 4 shows the distributions
$dF_2(x,Q^2,m_c^2,M)/dM$ and $dF_L(x,Q^2,m_c^2,M)/dM$
for various values of $x$ at $Q^2=12\, ({\rm GeV/c})^2$.
For charm production, the Born result multiplied by a constant
factor gives good agreement with the complete
${\cal O}(\alpha_s^2)$ result for the $2$ projection,
while the $L$ projection is not well reproduced for a
constant multiplicative factor for large invariant masses.
%While we have not shown the bottom distributions we note that
%they are very well reproduced by a constant multiplicative
%factor of $1.3$.
Figures 5 and 6 show the distributions $dF_2(x,Q^2,m_b^2,M)/dM$
and $dF_L(x,Q^2,m_b^2,M)/dM$ for bottom production
at $x=8.5 \times 10^{-4}$ while varying $Q^2$.
Figure 7 and 8 shows the distributions
$dF_2(x,Q^2,m_b^2,M)/dM$ and $dF_L(x,Q^2,m_b^2,M)/dM$
for bottom production
for various values of $x$ at $Q^2=12\, ({\rm GeV/c})^2$.
For bottom production, the Born result multiplied by a constant
factor gives quite good agreement with the complete
${\cal O}(\alpha_s^2)$ result for both the $2$ and $L$
projections.

To conclude, we repeat that the invariant mass distributions are
reasonably well represented by taking the Born result times a
multiplicative factor.  The agreement is excellent for bottom
production and not quite so good for charm production.

\vskip 0.5 cm
\centerline{\bf Acknowledgments}
We thank J. Whitmore for useful discussions and acknowledge S. Mendoza
for help in the early stages of the project.
Our research is supported in part by the contract NSF 9309888.

\newpage

\newcommand{\tema}{$\times 10^{-1}$}
\newcommand{\temb}{$\times 10^{-2}$}
\newcommand{\temc}{$\times 10^{-3}$}
\newcommand{\temd}{$\times 10^{-4}$}
\newcommand{\teme}{$\times 10^{-5}$}
\newcommand{\temf}{$\times 10^{-5}$}

\centerline{\bf \large{Table 1}}
\vspace{2cm}

\begin{center}
\begin{tabular}{||c|c||c|c|c||} \hline \hline
$x$ & $Q^2$ & $\mu=\mu_0/2$ & $\mu=\mu_0$ & $\mu=2\mu_0$ \\ \hline \hline
8.5 \temd & 8.5 & 7.36 \temb & 8.43 \temb & 9.12 \temb \\ \hline
8.5 \temd & 12  & 0.97 \tema & 1.10 \tema & 1.20 \tema \\ \hline
8.5 \temd & 25  & 1.57 \tema & 1.79 \tema & 1.94 \tema \\ \hline
8.5 \temd & 50  & 2.25 \tema & 2.52 \tema & 2.71 \tema \\ \hline
          &     &           &           &           \\ \hline
4.2 \temd & 12  & 1.16 \tema & 1.38 \tema & 1.54 \tema \\ \hline
8.5 \temd & 12  & 0.97 \tema & 1.10 \tema & 1.20 \tema \\ \hline
1.6 \temc & 12  & 8.03 \temb & 8.89 \temb & 9.38 \temb \\ \hline
2.7 \temc & 12  & 6.81 \temb & 7.29 \temb & 7.48 \temb \\ \hline
\hline
\end{tabular}
\end{center}
\vspace{2cm}

\begin{description}
\item[Table 1.]
Variation of $F_2$ with $\mu_0^2=Q^2+4(m_c^2+(P_{t_c}+P_{t_{\bar{c}}})^2/4)$
for various $x$ and $Q^2$ values.
\end{description}
\newpage

\centerline{\bf \large{Table 2}}
\vspace{2cm}

\begin{center}
\begin{tabular}{||c|c||c|c|c||} \hline \hline
$x$ & $Q^2$ & $\mu=\mu_0/2$ & $\mu=\mu_0$ & $\mu=2\mu_0$ \\ \hline \hline
8.5 \temd & 8.5 & 1.11 \temb & 1.23 \temb & 1.31 \temb \\ \hline
8.5 \temd & 12  & 1.68 \temb & 1.86 \temb & 1.99 \temb \\ \hline
8.5 \temd & 25  & 3.32 \temb & 3.64 \temb & 3.90 \temb \\ \hline
8.5 \temd & 50  & 5.07 \temb & 5.51 \temb & 5.86 \temb \\ \hline
          &     &           &           &           \\ \hline
4.2 \temd & 12  & 2.03 \temb & 2.33 \temb & 2.55 \temb \\ \hline
8.5 \temd & 12  & 1.68 \temb & 1.86 \temb & 1.99 \temb \\ \hline
1.6 \temc & 12  & 1.40 \temb & 1.50 \temb & 1.56 \temb \\ \hline
2.7 \temc & 12  & 1.18 \temb & 1.24 \temb & 1.26 \temb \\ \hline
\hline
\end{tabular}
\end{center}
\vspace{2cm}

\begin{description}
\item[Table 2.]
Variation of $F_L$ with $\mu_0^2=Q^2+4(m_c^2+(P_{t_c}+P_{t_{\bar{c}}})^2/4)$
for various $x$ and $Q^2$ values.
\end{description}
\newpage

\centerline{\bf \large{Table 3}}
\vspace{2cm}

\begin{center}
\begin{tabular}{||c|c||c|c|c||} \hline \hline
$x$ & $Q^2$ & $\mu=\mu_0/2$ & $\mu=\mu_0$ & $\mu=2\mu_0$ \\ \hline \hline
8.5 \temd & 8.5 & 1.53 \temc & 1.51 \temc & 1.47 \temc \\ \hline
8.5 \temd & 12  & 2.45 \temc & 2.45 \temc & 2.42 \temc \\ \hline
8.5 \temd & 25  & 6.28 \temc & 6.34 \temc & 6.36 \temc \\ \hline
8.5 \temd & 50  & 1.37 \temb & 1.40 \temb & 1.41 \temb \\ \hline
          &     &           &           &           \\ \hline
4.2 \temd & 12  & 3.37 \temc & 3.44 \temc & 3.47 \temc \\ \hline
8.5 \temd & 12  & 2.45 \temc & 2.45 \temc & 2.42 \temc \\ \hline
1.6 \temc & 12  & 1.77 \temc & 1.73 \temc & 1.66 \temc \\ \hline
2.7 \temc & 12  & 1.30 \temc & 1.24 \temc & 1.16 \temc \\ \hline
\hline
\end{tabular}
\end{center}
\vspace{2cm}

\begin{description}
\item[Table 3.]
Variation of $F_2$ with $\mu_0^2=Q^2+m_b^2+(P_{t_b}+P_{t_{\bar{b}}})^2/4$
for various $x$ and $Q^2$ values.
\end{description}
\newpage

\centerline{\bf \large{Table 4}}
\vspace{2cm}

\begin{center}
\begin{tabular}{||c|c||c|c|c||} \hline \hline
$x$ & $Q^2$ & $\mu=\mu_0/2$ & $\mu=\mu_0$ & $\mu=2\mu_0$ \\ \hline \hline
8.5 \temd & 8.5 & 5.34 \teme & 4.99 \teme & 4.52 \teme \\ \hline
8.5 \temd & 12  & 1.09 \temd & 1.04 \temd & 1.00 \temd \\ \hline
8.5 \temd & 25  & 4.65 \temd & 4.59 \temd & 4.50 \temd \\ \hline
8.5 \temd & 50  & 1.57 \temc & 1.56 \temc & 1.56 \temc \\ \hline
          &     &           &           &           \\ \hline
4.2 \temd & 12  & 1.44 \temd & 1.42 \temd & 1.39 \temd \\ \hline
8.5 \temd & 12  & 1.09 \temd & 1.04 \temd & 1.00 \temd \\ \hline
1.6 \temc & 12  & 8.21 \teme & 7.62 \teme & 7.08 \teme \\ \hline
2.7 \temc & 12  & 6.30 \teme & 5.67 \teme & 5.13 \teme \\ \hline
\hline
\end{tabular}
\end{center}
\vspace{2cm}

\begin{description}
\item[Table 4.]
Variation of $F_L$ with $\mu^2=Q^2+m_b^2+(P_{t_b}+P_{t_{\bar{b}}})^2/4$
for various $x$ and $Q^2$ values.
\end{description}
\newpage

%---------------- Figure Captions-----------------------------------------
% Figure Captions
\centerline{Figure Captions}
\begin{description}
\item[Fig.1.]
The distributions $dF_2(x,Q^2,m_c^2,M)/dM$ for charm production at
fixed $x=8.5 \times 10^{-4}$ and $Q^2=$ $8.5$ (solid line),
$12$ (dotted line), $25$ (short dashed line),
$50$ (long dashed line) all in units of $({\rm GeV/c})^2$.
Histograms are complete ${\cal O}( \alpha_s^2 )$ result.
For comparison we show the Born result times a multiplicative factor
of $1.3$ (empty box), $1.3$ (solid box), $1.2$ (empty circle), $1.2$
(solid circle) for the various $Q^2$ values, respectively.
\item[Fig.2.]
The distributions $dF_L(x,Q^2,m_c^2,M)/dM$ for charm production
at fixed $x=8.5 \times 10^{-4}$ while varying $Q^2$.
Notation is that of Figure 1 for the ${\cal O}( \alpha_s^2 )$ result.
For comparison we show the Born result times a multiplicative factor
of $1.9$ (empty box), $1.9$ (solid box), $1.6$ (empty circle), $1.6$
(solid circle) for the various $Q^2$ values, respectively.
\item[Fig.3]
The distributions $dF_2(x,Q^2,m_c^2,M)/dM$ for charm production
at fixed $Q^2=12 \, ({\rm GeV/c})^2$ with $x=$ $4.2 \times 10^{-4}$
(solid line), $8.5 \times 10^{-4}$ (dotted line), $1.6 \times 10^{-3}$
(short dashed line), $2.7 \times 10^{-3}$ (long dashed line).
For comparison we show the Born result times a multiplicative factor
of $1.3$ (empty box), $1.3$ (solid box), $1.3$ (empty circle), $1.4$
(solid circle) for the various $x$ values, respectively.
\item[Fig.4]
The distributions $dF_L(x,Q^2,m_c^2,M)/dM$ for charm production
at fixed $Q^2=12 \, ({\rm GeV/c})^2$ while varying $x$.
Notation is that of Figure 3 for the ${\cal O}( \alpha_s^2 )$ result.
For comparison we show the Born result times a multiplicative factor
of $1.9$ (empty box), $1.9$ (solid box), $1.9$ (empty circle), $1.5$
(solid circle) for the various $x$ values, respectively.
\item[Fig.5.]
The distributions $dF_2(x,Q^2,m_b^2,M)/dM$ for bottom production
at fixed $x=8.5 \times 10^{-4}$ while varying $Q^2$. Notation is
that of Figure 1 for the ${\cal O}( \alpha_s^2 )$ result
but all points are $1.3 \times$ Born result
for the various $Q^2$ values, respectively.
\item[Fig.6.]
The distributions $dF_L(x,Q^2,m_b^2,M)/dM$  for bottom production
at fixed $x=8.5 \times 10^{-4}$ while varying $Q^2$. Notation is
that of Figure 1 for the ${\cal O}( \alpha_s^2 )$ result
but all points are $1.3 \times$ Born result
for the various $Q^2$ values, respectively.
\item[Fig.7]
The distributions $dF_2(x,Q^2,m_b^2,M)/dM$ for bottom production
while varying $x$ at fixed $Q^2=12 \, ({\rm GeV/c})^2$. Notation is
that of Figure 2 for the ${\cal O}( \alpha_s^2 )$ result
but all points are $1.3 \times$ Born result
for the various $x$ values, respectively.
\item[Fig.8]
The distributions $dF_L(x,Q^2,m_b^2,M)/dM$ for bottom production
while varying $x$ at fixed $Q^2=12 \, ({\rm GeV/c})^2$. Notation is
that of Figure 2 for the ${\cal O}( \alpha_s^2 )$ result
but all points are $1.3 \times$ Born result
for the various $x$ values, respectively.
\end{description}

\newpage

%
%----------------------------References-------------------------------------
%
%%\begin{references}

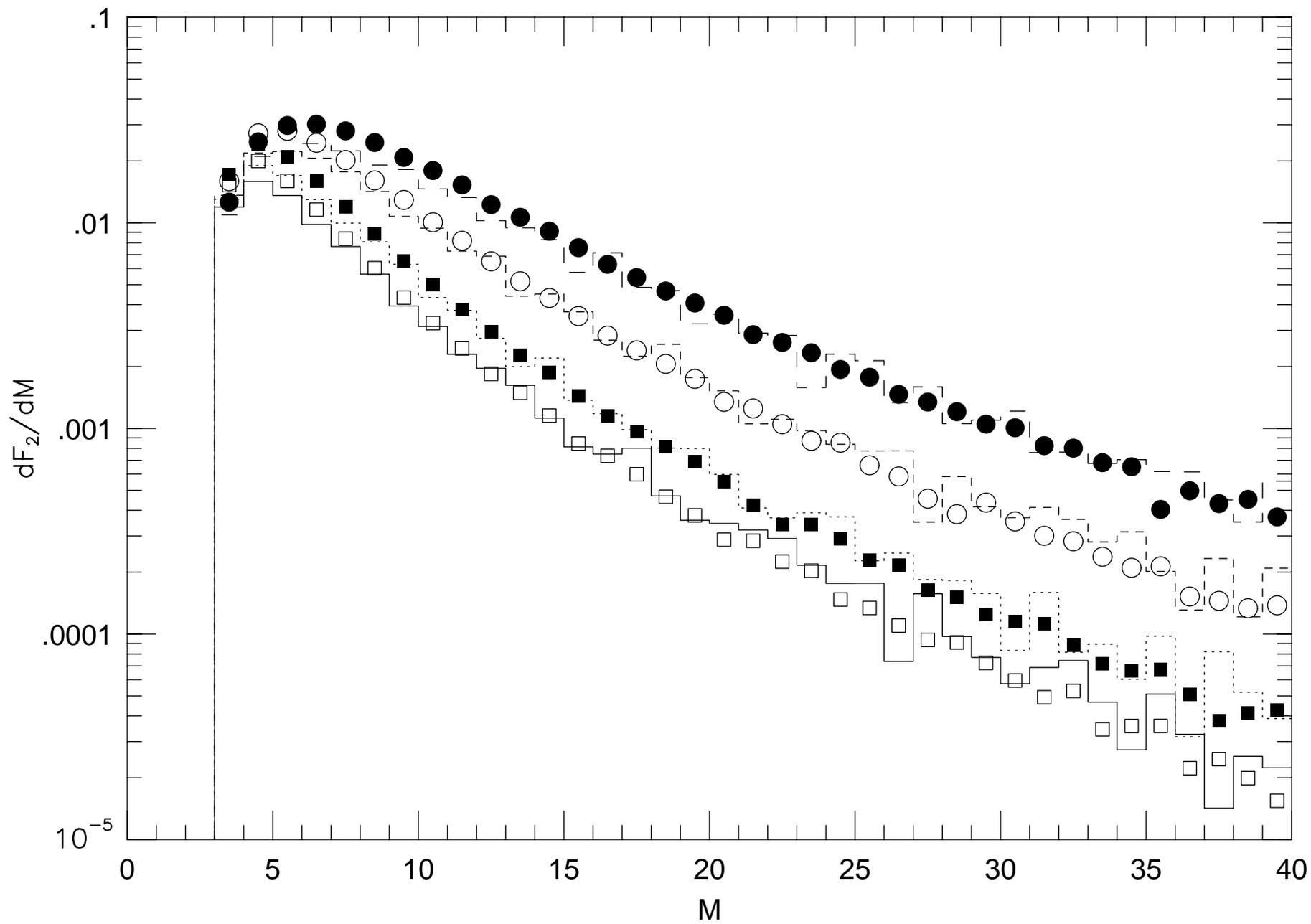
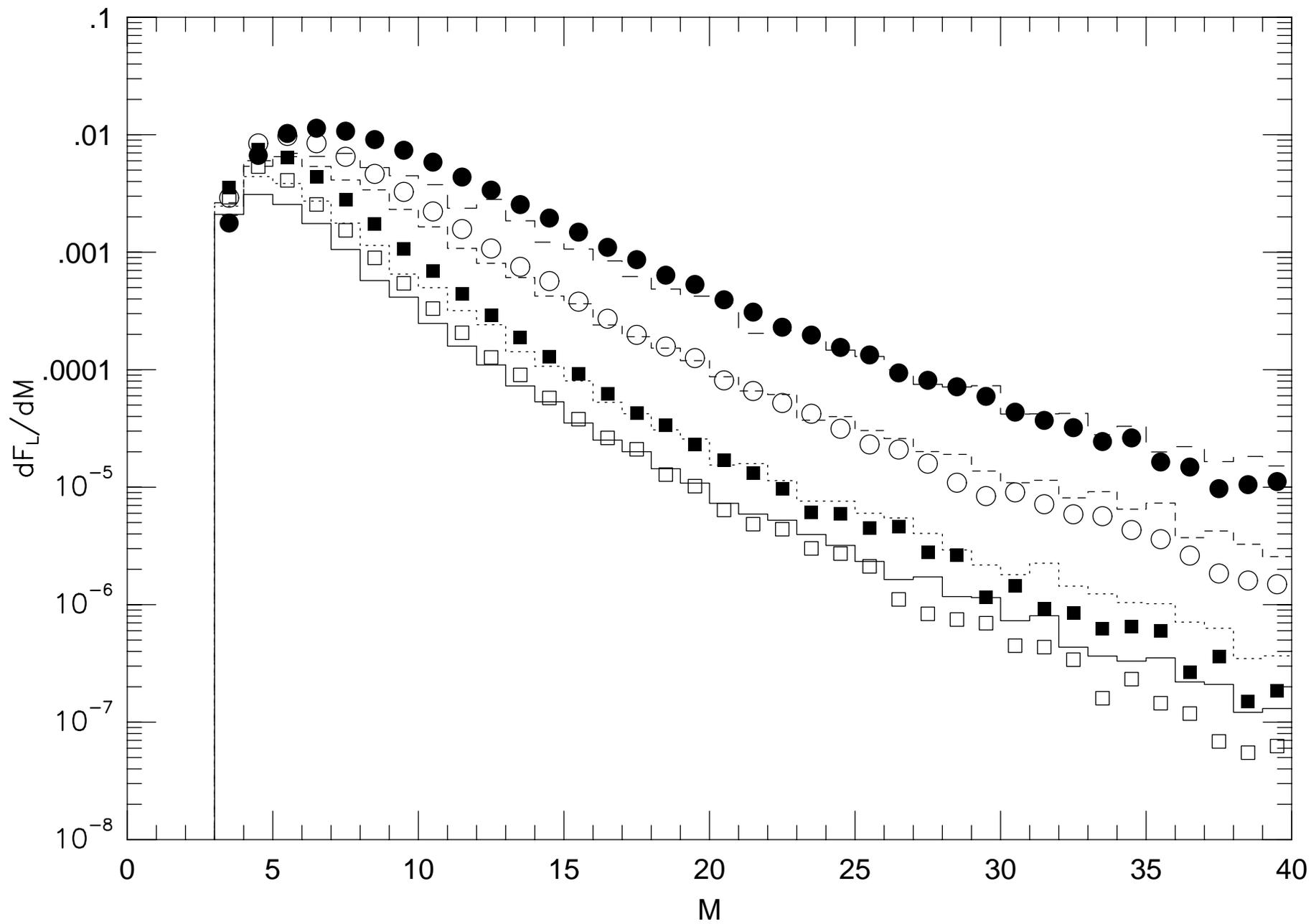
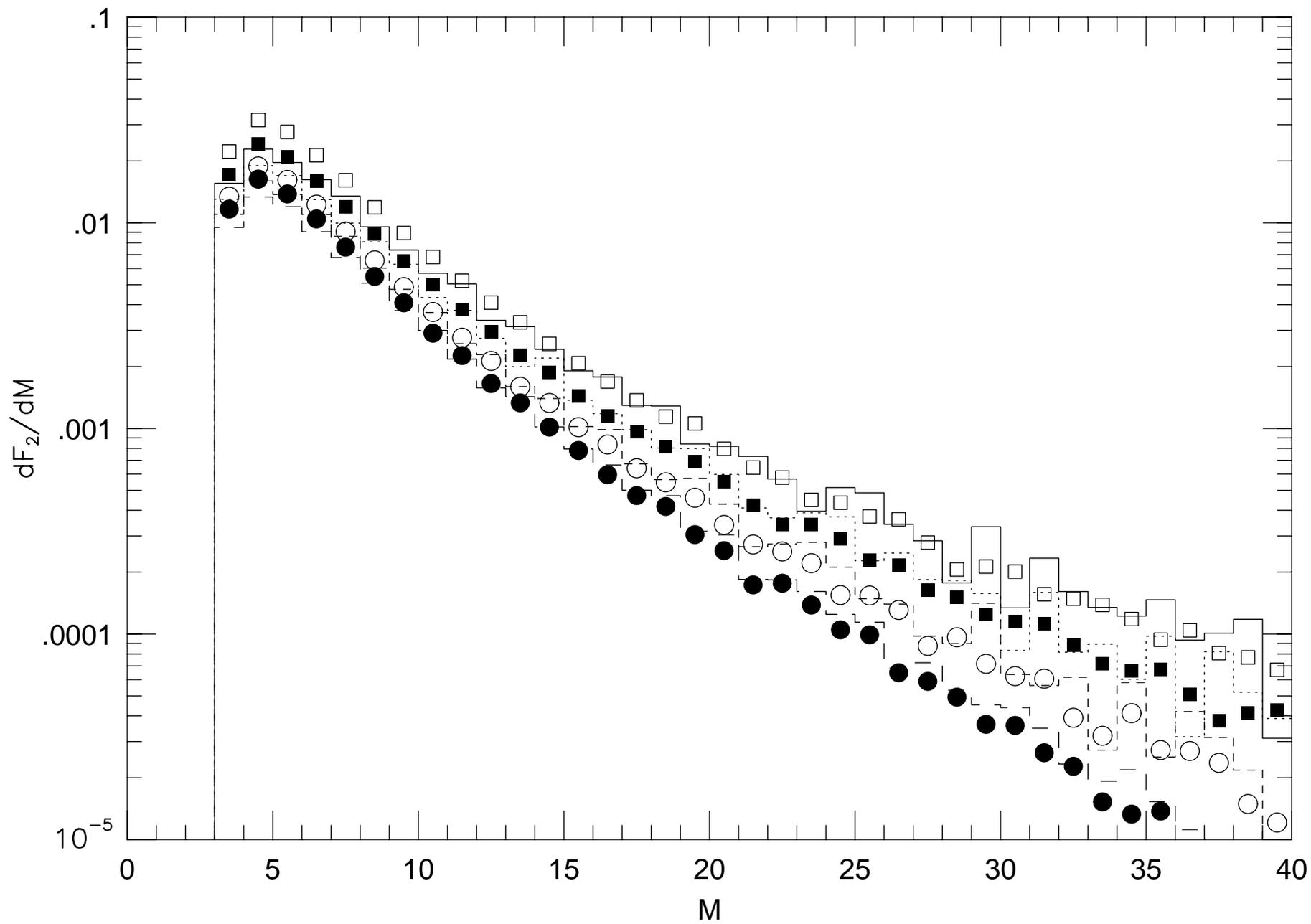
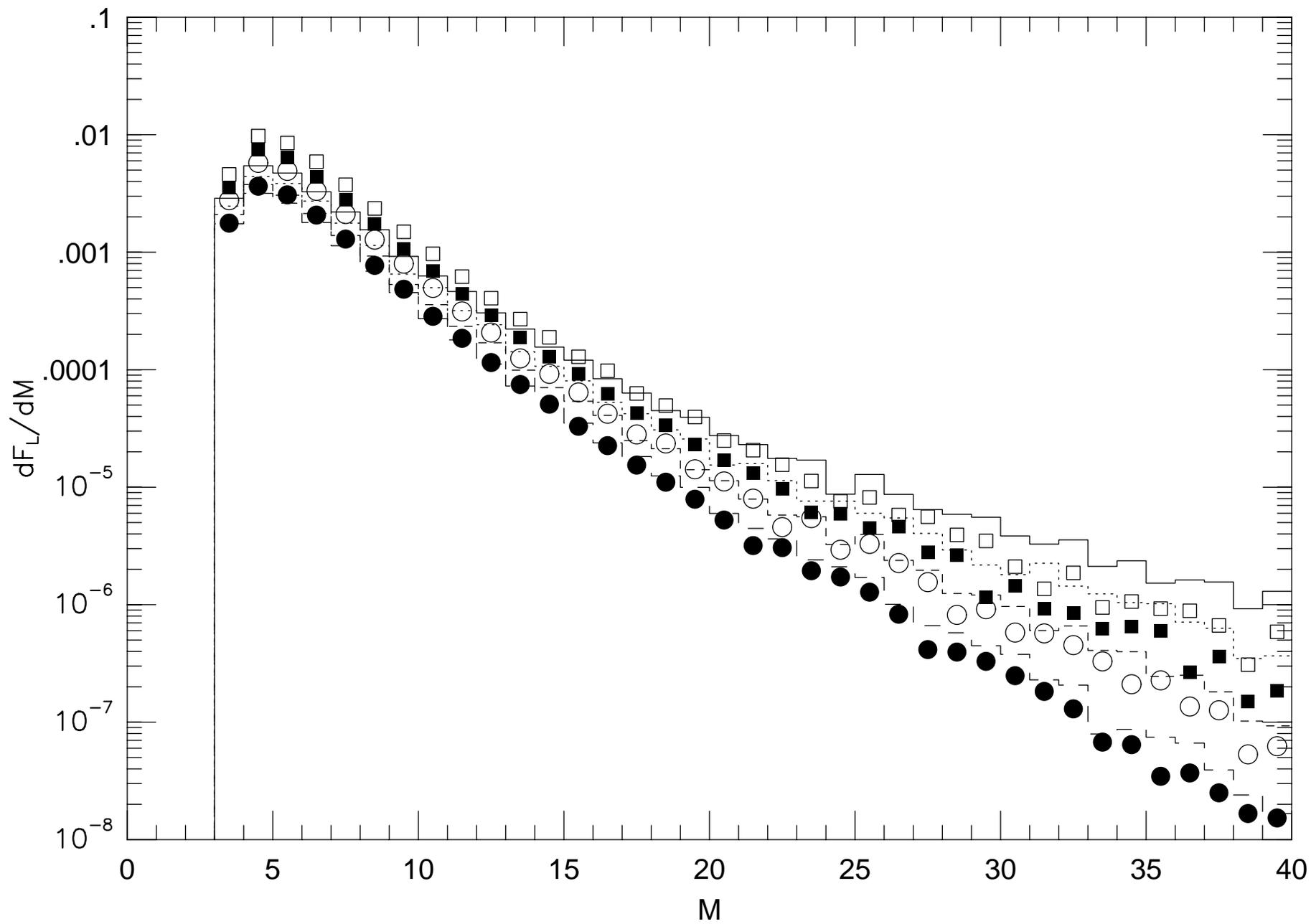
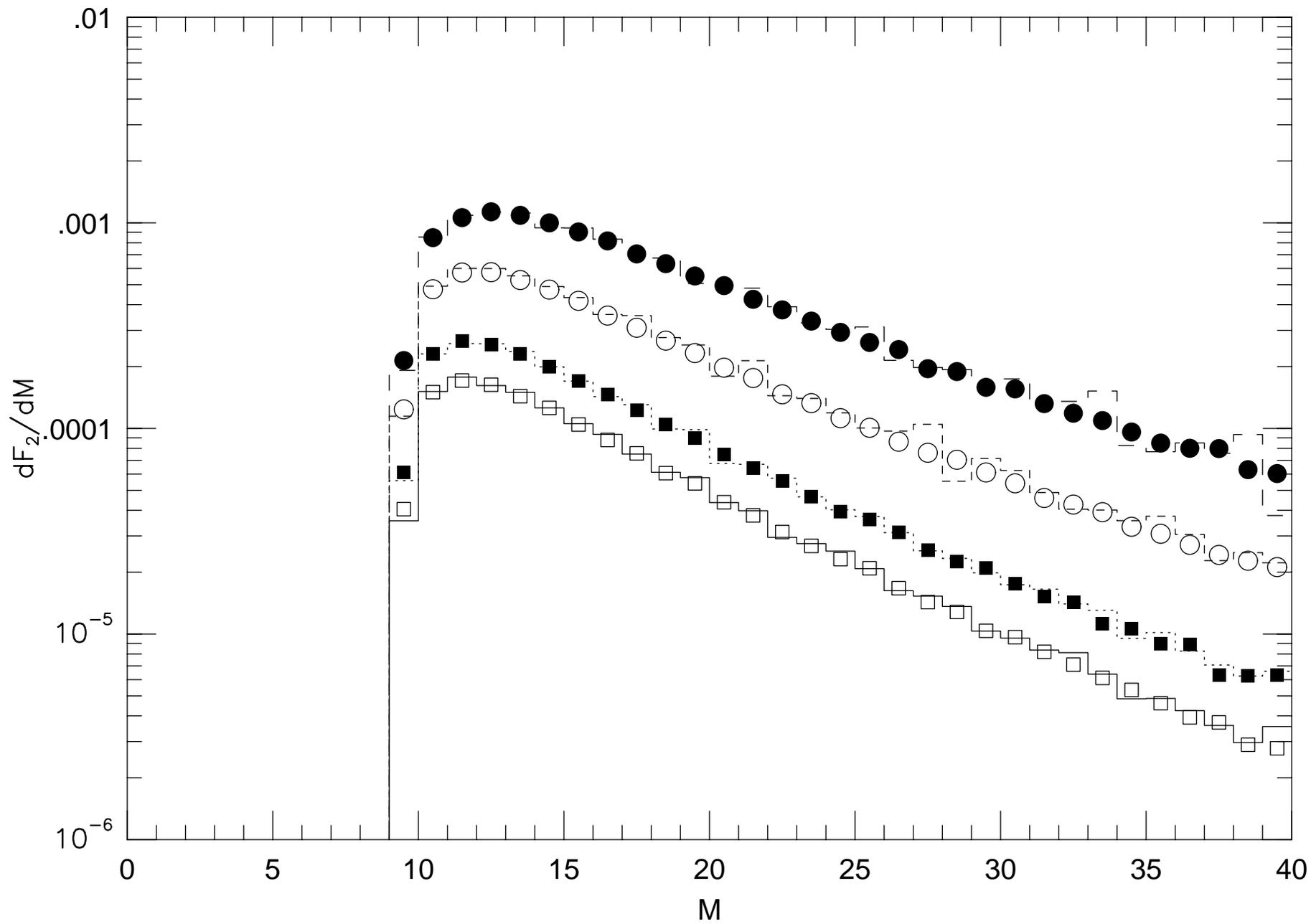
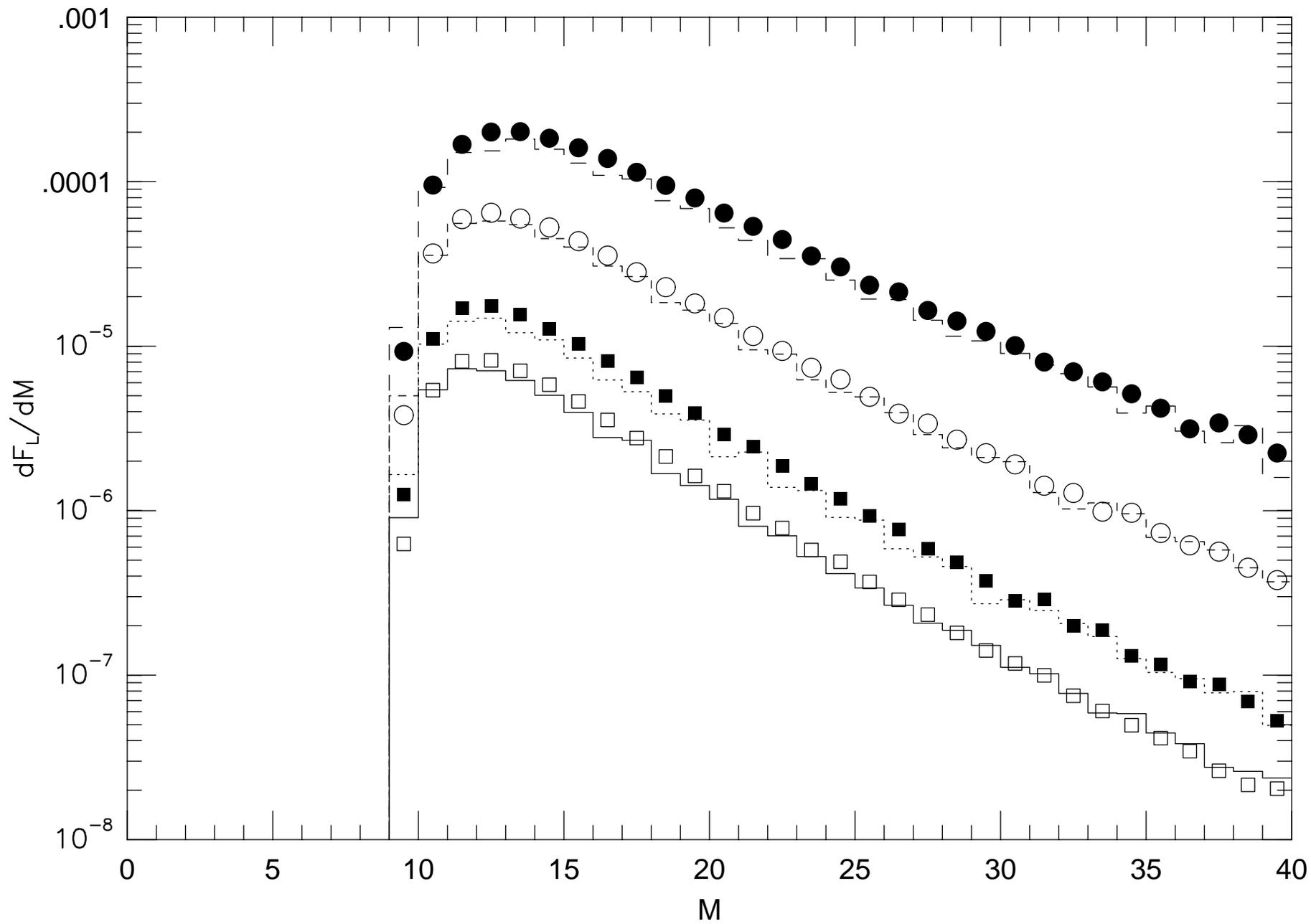
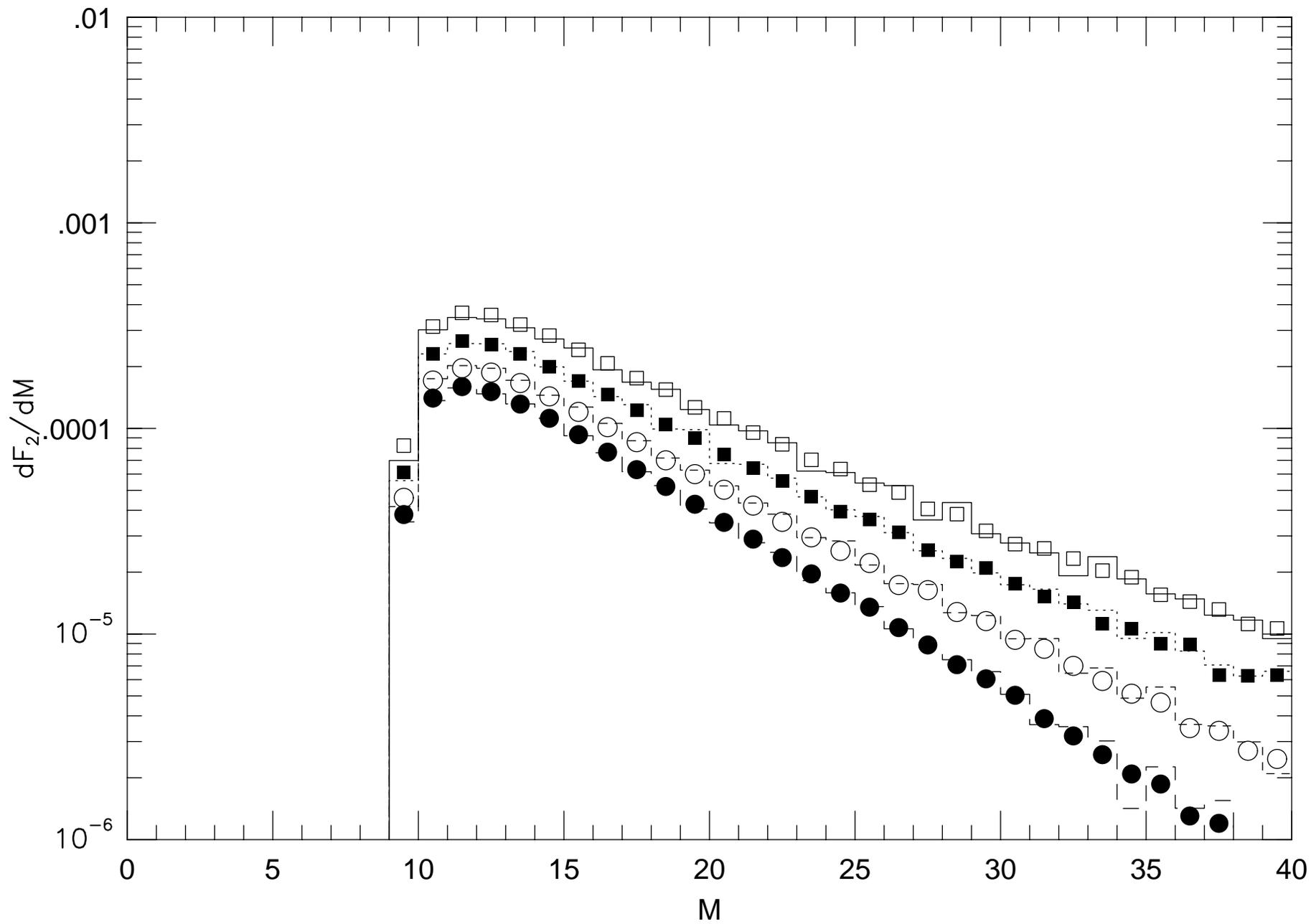
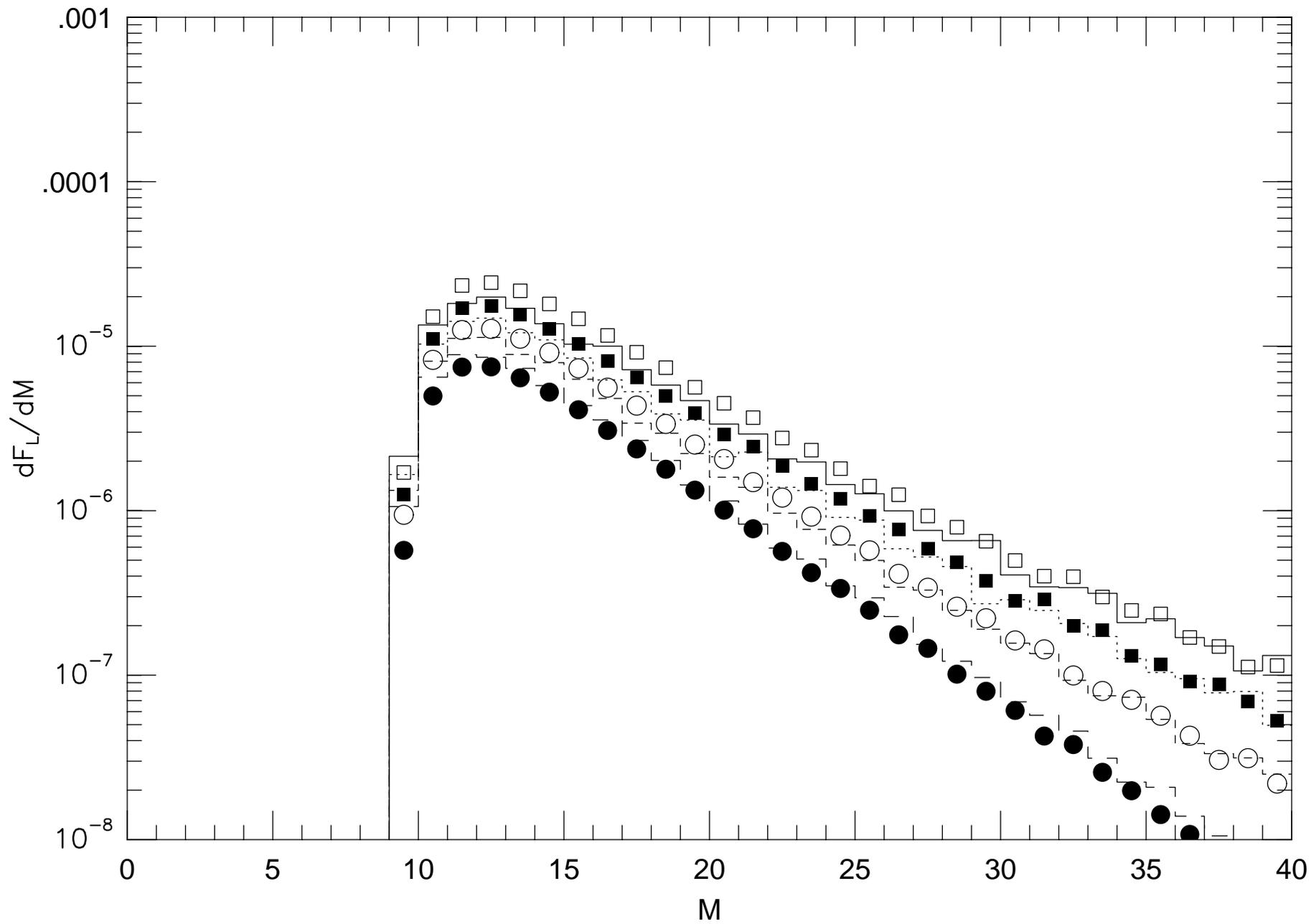
\end{document}